\documentclass[fleqn]{2017SCGE}

\begin{document}

\ensubject{subject}
\ArticleType{Article}
\SpecialTopic{SPECIAL TOPIC: }
\Year{}
\Month{}
\Vol{}
\No{}
\DOI{}
\ArtNo{000000}
\ReceiveDate{}
\AcceptDate{}

\title{Quantum-enhanced interferometry with asymmetric beam splitters}

\author[1,2]{Wei Zhong}{}
\author[1]{Fan Wang}{}
\author[3]{Lan Zhou}{}
\author[1,2]{Peng Xu}{}
\author[1,4,]{Yu-Bo Sheng}{{shengyb@njupt.edu.cn}}
\AuthorMark{Zhong W}

\AuthorCitation{Zhong W, Wang F, Zhou L, Xu P, Sheng Y B}

\address[1]{Institute of Quantum Information and Technology, Nanjing University
of Posts and Telecommunications, Nanjing 210003, China}
\address[2]{National Laboratory of Solid State Microstructures, Nanjing University,
Nanjing 210093, China}
\address[3]{School of Science, Nanjing University of Posts and Telecommunications,
Nanjing 210003, China}
\address[4]{Key Lab of Broadband Wireless Communication and Sensor Network Technology,
Nanjing University of Posts and Telecommunications, \\Ministry of Education,
Nanjing 210003, China}

\abstract{In this paper, we investigate the phase sensitivities in two-path optical interferometry with asymmetric beam splitters. Here, we present the optimal conditions for the transmission ratio and the phase of the beam splitter to gain the highest sensitivities for a general class of non-classical states with parity symmetry. Additionally, we address the controversial question of whether the scheme with a combination of coherent state and photon-added or photon-subtracted squeezed vacuum state is better or worse than the most celebrated one using a combination of coherent state and squeezed vacuum state.}


\maketitle
\begin{multicols}{2}

\section{Introduction}

As a valuable tool in precision measurement, optical interferometer
has been used for practical applications, such as the measurement of the speed of light, microscopic imaging, and detection of gravitational waves. Classically, the ultimate sensitivity of this tool is limited by the so-called shot noise limit (SNL), due to quantum fluctuation of classical light of use. It was shown that this limit can be beaten by using non-classical states
\cite{Caves1981PRD,Holland1993PRL,Luis2001PRA,Dorner2009PRL,Anisimov2010PRL,Joo2011PRL,Pezze2013PRL,Tan2014PRA,Lang2014PRA,Birrittella2014JOSAB,Yuan2014,Zhang2016,Ouyang2016JOSAB,Liu2019}.
Because of this, there have been enormous efforts made to develop probe strategies by using quantum states, which aimed to reach the so-called Heisenberg limit (HL) \cite{Rarity1990PRL,Nie2018}. Recently, several works have found some optimal feasible detection schemes to access the ultimate sensitivities for the abovementioned strategies \cite{Pezze2008PRL,Hofmann2009PRA,Krischek2011PRL,Seshadreesan2013PRA,Pezze2013PRL,Lang2013PRL,Vidrighin2014NC,Zhong2017PRA,Liu2017PRA}.

A generic phase measurement procedure consists of three parts: probes, phase accumulation, and detection \cite{Helstrom1976Book,Holevo1982Book,Braunstein1994PRL}. Besides optimizing the probe state and detection, we can further enhance the phase sensitivity by optimizing the phase accumulation. As shown in Fig.~\ref{fig:MZI}, the phase accumulation process in a linear optical interferometry \cite{Long2018,Qin2019} consists of beam splitting, phase shift (PS), and beam emerging, in which the splitting and emerging are modeled by two beam splitters (BSs) quantified by two parameters: transmission ratio and phase. Since the PS is given, the optimization of phase accumulation is then equivalent to the optimization of the transmission coefficient of the BS. It is common to select a balanced BS with a fixed phase of $0$ or $\pm\pi/2$. However, there are no studies that report whether such a choice is optimal for a generic state (except for special cases \cite{Jarzyna2012PRA}) and what are the optimal conditions for both transmission ratio and phase of the BS to gain the highest phase sensitivity.

In this work, we address these issues by explicitly investigating the ultimate sensitivity of interferometry with asymmetric BSs according to the single- and multi-parameter estimation theories, respectively. We consider the probe state to be a general family of quantum states with parity symmetry. This family covers a wide range of states employed in most of current interferometric phase measurements \cite{Caves1981PRD,Holland1993PRL,Anisimov2010PRL,Luis2001PRA,Joo2011PRL,Pezze2013PRL,Tan2014PRA,Birrittella2014JOSAB,Lang2014PRA,Ouyang2016JOSAB}.
Then, we demonstrate the optimal conditions for the transmission ratio and the phase of the BS by maximizing the quantum Fisher information (QFI). We apply these results to various typical state strategies and analyze their sensitivities given the constraint on the total photon number. Based on the multi-parameter estimation theory, we find that a balanced BS is optimal for most of states of consideration and the phase of the BS is strictly relevant to the typical state of use. According to single-parameter estimation theory, it is shown that a balanced BS may be the worst option under certain conditions. In addition, we revisit the interferometric schemes by using a combination of coherent state and photon-added and photon-subtracted squeezed vacuum state (CS$\otimes$PASVS and CS$\otimes$PSSVS). It has been found that these schemes may outperform the celebrated one of using a mixing of coherent state with squeezed vacuum state (CS$\otimes$SVS), which seems to contradict the conclusion made by Lang and Caves \cite{Lang2013PRL}. It remains ambiguous whether CS$\otimes$PASVS and CS$\otimes$PSSVS may outperform CS$\otimes$SVS for the phase sensitivity. We clarify this by analytically calculating the QFI.

This paper is organized as follows. In Sec. II, we briefly introduce the setup of Mach-Zehnder interferometer (MZI) and obtain the ultimate sensitivities with asymmetric BS. In Sec. III, we present the optimal conditions for the transmission coefficient of the BS and apply them to specific quantum states. A further discussion of the sensitivity enhanced by CS$\otimes$PASVS and CS$\otimes$PSSVS has been made in Sec. IV. Finally, our conclusions are presented in Sec. V.

\section{Ultimate phase sensitivity in MZI with asymmetric beam splitter}

\begin{figure}[H]
\centering
\includegraphics[width=1\columnwidth]{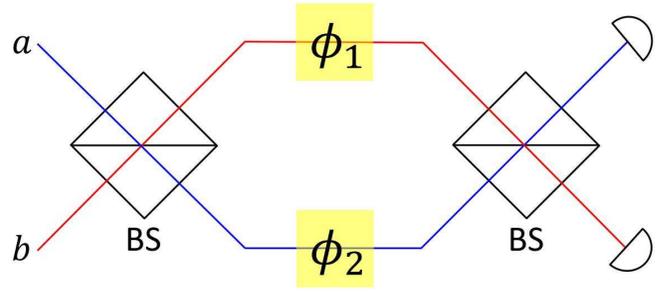}
\caption{(Color online) Schematic of the MZI consisting of two BSs and a PS.}
\label{fig:MZI}
\end{figure}

MZI is modeled as a two-mode linear optical interferometer with $a$ and $b$ denoting annihilation operators of upper and lower input modes, respectively (see Fig.~\ref{fig:MZI}). In general, the MZI setup is made up of two BSs denoted by $B$ and a phase shift denoted by $U$. Thus, the total transformation of the MZI is represented as a compound operation of $B^{\dagger}UB$. Let $\vert\psi_{{\rm in}}\rangle$ denote the state entering at the input ports of the interferometer. Then, the state at the output ports reads $\vert\psi_{{\rm out}}\rangle=B^{\dagger}UB\vert\psi_{{\rm in}}\rangle$.

More generally, BS and PS operations can be explicitly modeled, respectively, by \cite{Yurke1986PRA}
\begin{eqnarray}
B & = & \exp\left[-i\left(\gamma a^{\dagger}b+\gamma^{\ast}ab^{\dagger}\right)/2\right],\label{eq:BS0}\\
U & = & \exp\left[-i\left(\phi_{2}a^{\dagger}a+\phi_{1}b^{\dagger}b\right)\right],\label{eq:PS0}
\end{eqnarray}
with the complex quantity $\gamma=\tau e^{i\vartheta}$ where the modulation $\tau$ relates to the transmission ratio $T=\cos^{2}\frac{\tau}{2}$ and the argument $\vartheta$ denotes the phase. $\phi_{1}$ and $\phi_{2}$ are unknown phase parameters imprinted on each arm of the interferometer. It is worth to note that we choose here $\tau$ to quantify the transmission ratio, rather than $T$, since it will facilitate our analysis below. By using $B$ of Eq.~(\ref{eq:BS0}), one can easily obtain the desired input-output relation of a generic
BS
\begin{eqnarray}
B^{\dagger}\left(\begin{array}{c}
a\\
b
\end{array}\right)B & = & \left(\begin{array}{c}
a\cos\frac{\tau}{2}-ie^{i\vartheta}b\sin\frac{\tau}{2}\\
b\cos\frac{\tau}{2}-ie^{-i\vartheta}a\sin\frac{\tau}{2}
\end{array}\right).
\end{eqnarray}

A routine means for studying MZI is to use the Schwinger representation
\begin{eqnarray}
J_{x}=\frac{1}{2}\left(a^{\dagger}b+ab^{\dagger}\right), & \quad & J_{y}=\frac{1}{2i}\left(a^{\dagger}b-ab^{\dagger}\right),\\
J_{z}=\frac{1}{2}\left(a^{\dagger}a-b^{\dagger}b\right).
\end{eqnarray}
They satisfy the commutation relations for Lie algebra of$\mathfrak{su}\left(2\right)$
\begin{eqnarray}
[J_{x},\thinspace J_{y}]=iJ_{z}, & \quad  [J_{y},\thinspace J_{z}]=iJ_{x},& \quad[J_{z},\thinspace J_{x}]=iJ_{y}.
\end{eqnarray}
and commute with the rescaled total photon number operator
\begin{eqnarray}
\normalcolor J_{0}&=&\frac{N}{2}=\frac{1}{2}\left(a^{\dagger}a+b^{\dagger}b\right),
\end{eqnarray}
i.e., $[J_{i},\thinspace J_{0}]=0,\thinspace\left(i=x,y,z\right)$.
Based on this representation, one can rewrite Eq.~(\ref{eq:BS0})
and Eq.~(\ref{eq:PS0}) as follows:

\begin{eqnarray}
B & = & \exp\left[-i\tau\left(\cos\vartheta J_{x}+\sin\vartheta J_{y}\right)\right],\label{eq:BS}\\
U & = & \exp\left[-i\left(\phi_{s}J_{0}+\phi_{d}J_{z}\right)\right],\label{eq:PS}
\end{eqnarray}
where we denote the sum and difference PSs as $\phi_{s}=\phi_{1}+\phi_{2}$
and $\phi_{d}=\phi_{2}-\phi_{1}$. Obviously, from Eq.~(\ref{eq:BS}),
$\tau=\pi/2$ (i.e., $T=1/2$) corresponds to a $50:50$ BS, and $\vartheta=0\left(-\pi/2\right)$
corresponds to a clockwise rotation operation of $\pi/2$ along $x\left(y\right)$ axis. These typical values of $\tau$ and $\vartheta$ were commonly chosen in previous studies. Optimization of these two quantities shall be made by maximizing the QFI of a broad class of states.

Quantum estimation theory states that the phase sensitivity is theoretically limited by the inverse of the QFI (see below for detailed definition of the QFI), which is only dependent on the parametric state \cite{Braunstein1994PRL}. It means that the larger value of the QFI is the higher sensitivity of phase estimation could acquire. To get the theoretical sensitivity limit, one should apply optimal measurements in practice, which have been widely investigated \cite{Braunstein1994PRL,Pezze2008PRL,Lang2013PRL,Seshadreesan2013PRA,Vidrighin2014NC,Hofmann2009PRA,Krischek2011PRL,Zhong2017PRA}.
In our case, the parametric state is given by $\vert\psi_{{\rm out}}\rangle=B^{\dagger}UB\vert\psi_{{\rm in}}\rangle$. From Eq.~(\ref{eq:BS}), the BS does not depend on the value of the phase shift to be estimated. Thus, the QFI of $\vert\psi_{{\rm out}}\rangle$ is equivalent to that of $UB\vert\psi_{{\rm in}}\rangle$, as a consequence of the property that the QFI is invariant under the parameter-independent unitary operation \cite{Lu2010PRA,Zhong2013PRA,Liu2019JPA}. For given $\vert\psi_{{\rm in}}\rangle$ and $U$,
maximizing the QFI reduces to optimization on the first BS.

According to the quantum Cram\'er-Rao theorem \cite{Helstrom1976Book,Holevo1982Book,Braunstein1994PRL}, the sensitivity of simultaneously estimating multi-parameter is measured by the covariance matrix of the estimators denoted by $\Sigma$, and the sensitivity of the unbiased estimators is limited by the inverse of the QFI matrix $F$, i.e., $\Sigma\geq\left(\upsilon F\right)^{-1}$, up to a number of independent measurements $\upsilon$. In our case, the estimators are denoted by $\phi_{s}$ and $\phi_{d}$. Then the quantum Cram\'er-Rao inequality can be simply represented as
\begin{equation}
\left(\begin{array}{cc}
\mathcal{V}(\hat{\phi}_{s}) & \mathcal{C}(\hat{\phi}_{s},\hat{\phi}_{d})\\
\mathcal{C}(\hat{\phi}_{s},\hat{\phi}_{d}) & \mathcal{V}(\hat{\phi}_{d})
\end{array}\right)\geq\frac{1}{\upsilon(F_{ss}F_{dd}-F_{sd}^{2})}\left(\begin{array}{cc}
F_{dd} & -F_{sd}\\
-F_{sd} & F_{ss}
\end{array}\right),\label{eq:two-QCRB}
\end{equation}
where $\mathcal{V}$ and $\mathcal{C}$ represent the variance and the covariance, respectively, and the inverse of the QFI matrix $F$ is in terms of three elements
\begin{eqnarray}
F_{ss} & = & 4\mathcal{V}(J_{0}),\label{eq:Fss}\\
F_{sd} & = & 4\mathcal{C}(J_{0},B^{\dagger}J_{z}B),\label{eq:Fsd}\\
F_{dd} & = & 4\mathcal{V}(B^{\dagger}J_{z}B),\label{eq:Fdd}
\end{eqnarray}
with
\begin{eqnarray}
B^{\dagger}J_{z}B & = & J_{z}\cos\tau+\left(J_{y}\cos\vartheta+J_{x}\sin\vartheta\right)\sin\tau.\label{eq:Jz-B}
\end{eqnarray}
See Appendix A for explicit expressions of $F_{ss}$, $F_{dd}$ and $F_{sd}$ with respect to mode operators $a$ and $b$. Here, the variance and the covariance of operators are separately denoted as $\mathcal{V}\left(\mathcal{O}\right)\equiv\langle\left(\Delta\mathcal{O}\right)^{2}\rangle$
and $\mathcal{C}\left(\mathcal{P},\mathcal{Q}\right)\equiv\frac{1}{2}\langle\left[\mathcal{P},\mathcal{Q}\right]_{+}\rangle-\left\langle \mathcal{P}\right\rangle \left\langle \mathcal{Q}\right\rangle $
with $\left[\bullet,\bullet\right]_{+}$ denoting an anti-commutator. Note that above, the commutation relationship of $\left[J_{0},B\right]=0$ has been considered, and the expectation $\left\langle \bullet\right\rangle $ here and below is defined on the input state $\vert\psi_{{\rm in}}\rangle$. It is worth to note that the above two-parameter estimation scenario (given by Eq.~(\ref{eq:two-QCRB})) reduces to two independent single-parameter cases when ${\color{blue}{\normalcolor F_{sd}=0}}$
\cite{Takeoka2017PRA}.

\begin{table*}[!]
\centering
\caption{The ultimate phase sensitivities for different input states, including TFS, CS$\otimes$FS, CS$\otimes$CSS, CS$\otimes$SVS, TSVS, and TMSVS. Correspondingly, the optimal transmission ratio $\tau$ and the phase $\vartheta$ of the BS are listed based on two- and single-parameter estimation theories depicted by Eqs.~(\ref{eq:two-QCRB-phid}) and (\ref{eq:single-QCRB-phid}). In this table, we find that both sensitivity bounds of Eqs.~(\ref{eq:two-QCRB-phid}) based on the two-parameter estimation theory and (\ref{eq:single-QCRB-phid}) based on the single-parameter estimation theory give the same $\tau_{{\rm opt}}$ and $\vartheta_{{\rm opt}}$ for different input states, except for the case of CS$\otimes$SVS. For CS$\otimes$SVS, $\tau_{{\rm opt}}$ and $\vartheta_{{\rm opt}}$ listed in the table are obtained from two-parameter estimation theory, and the optimal BS parameters based on single-parameter estimation theory are clarified in the table note.  Here, we set $\alpha=\left|\alpha\right|e^{i\varphi_{\alpha}}$, $\beta=\left|\beta\right|e^{i\varphi_{\beta}}$,
and $\xi=\left|\xi\right|e^{i\varphi_{\xi}}$. \label{tab:QFI}}

\resizebox{1\textwidth}{70pt}{
\begin{tabular}{cccccc}
\hline
\hline
Input States & $\mathfrak{G}$ & $4\mathcal{V}(J_{z})$ & $\mathfrak{F}$ & $\tau_{\rm{opt}}$ & $\vartheta_{\rm{opt}}$ \\

\hline

TFS & $0$ & $0$ & $2\left(n_{a}^{2}+n_{a}\right)$ & $\pi/2$ & $\left[-\pi,\pi\right]$ \\

CS$\otimes$FS & $0$ & $n_{a}$ & $2n_{a}n_{b}+n_{a}+n_{b}$ & $\pi/2$ & $\left[-\pi,\pi\right]$ \\

CS$\otimes$CSS & $\frac{4n_{a}n_{b}}{n_{a}+n_{b}}$ & $n_{a}+2(n_{b}^{2}+n_{b}$ & $4n_{a}n_{b}+n_{a}+n_{b}$ & $\pi/2$ & $(2\varphi_{\alpha}\!-2\varphi_{\beta}\pm\pi)/2$ \\

CS$\otimes$SVS & $\frac{8n_{a}\left(n_{b}^{2}+n_{b}\right)}{n_{a}+2\left(n_{b}^{2}+n_{b}\right)}$ & $n_{a}+2(n_{b}^{2}+n_{b})$ & $2n_{a}n_{b}+n_{a}+n_{b}+2n_{a}\sqrt{n_{b}^{2}+n_{b}}$ & $\pi/2\;\tnote{$\ast$}$ & $(2\varphi_{\alpha}-\varphi_{\xi})/2\;(\ast)$ \\

TSVS & $4\left(n_{a}^{2}+n_{a}\right)$ & $4\left(n_{a}^{2}+n_{a}\right)$ & $4\left(n_{a}^{2}+n_{a}\right)$ & $\left[0,\pi/2\right]$ & $\pm\pi/2$ \\

TMSVS & / & $0$ & $4\left(n_{a}^{2}+n_{a}\right)$ & $\pi/2$ & $\left[-\pi,\pi\right]$\\
\hline
\hline
\end{tabular}}
\begin{tablenotes}
\footnotesize

\item[] $\ast$ The optimal $\tau_{\rm{opt}}$ and $\vartheta_{\rm{opt}}$ listed here for $\mathcal{V}_{1}(\hat{\phi}_{d})$ based on single-parameter estimation theory only hold under the condition of $n_{b}<2n_{a}$, while when $n_{b}\geq2n_{a}$, $\mathcal{V}_{1}(\hat{\phi}_{d})$ gets the minimum at $\tau_{\rm{opt}}=0$ and $\vartheta_{\rm{opt}}\in[-\pi,\pi]$ due to $4\mathcal{V}(J_{z})>\mathfrak{F}$.

\end{tablenotes}
\centering
\end{table*}

Unlike the case of single-parameter estimation, the sensitivity bound for multi-parameter cases may not be always saturated \cite{Matsumoto2002JPA}, but it can be asymptotically reached when the generators of each parameter commute to each other \cite{Baumgratz2016PRL,Gagatsos2016PRA,Pezze2017PRL}. In MZI, the sum phase $\phi_{s}$ cannot be generally resolved with rare photon-counting detection without the introduction of additional resources (such as a homodyne measurement). That is because the resolution of $\phi_{s}$ depends on coherences between states with different numbers of photon, and these coherences can be interfered with an additional reference beam \cite{Molmer1997PRA,Demkowicz-Dobrzanski2009PRA,Jarzyna2012PRA}. Thus, the difference phase parameter $\phi_{d}$ is of more interest with respect to the situation in the absence of reference beam. From inequality~(\ref{eq:two-QCRB}), the ultimate estimation sensitivity of $\phi_{d}$ is bounded by
\begin{eqnarray}
\mathcal{V}_{2}(\hat{\phi}_{d}) & \geq & \frac{1}{\upsilon}\frac{F_{ss}}{F_{ss}F_{dd}-F_{sd}^{2}}.\label{eq:two-QCRB-phid}
\end{eqnarray}
This inequality is general for interferometric phase measurement.
Obviously, when $F_{sd}=0$, Eq.~(\ref{eq:two-QCRB-phid}) reduces
to
\begin{eqnarray}
\mathcal{V}_{1}(\hat{\phi}_{d})&\geq&\left(\upsilon F_{dd}\right)^{-1},\label{eq:single-QCRB-phid}
\end{eqnarray}
which is the quantum Cram\'er-Rao inequality for single-parameter estimation that is widely used in previous studies. In above, the subscripts ``$1$'' and ``$2$'' have been added to the $\mathcal{V}(\hat{\phi}_{d})$ to emphasize that the sensitivity bounds are obtained from single- and two-parameter quantum Cram\'er-Rao inequalities, respectively.

Furthermore, we should present the extent of applications of the two sensitivity bounds given above. In some cases, one does not want to know each parameter in a multi-parameter estimation. These unwanted parameters are called nuisance parameters \cite{Trees2013Book}. In our interferometric case, the sum phase serves as the nuisance parameter, while the difference phase is of most interest.  If the nuisance phase $\phi_{s}$ is unknown, then the sensitivity should be limited by the lower bound given by Eq.~(\ref{eq:two-QCRB-phid}). This means that the unknown nuisance parameter $\phi_{s}$ affects the sensitivity of estimation of the wanted parameter $\phi_{d}$. If the nuisance phase $\phi_{s}$ is accurately known, then the above two-phase estimation problem belongs to the single-parameter estimation, in which the sensitivity bound should be described by Eq.~(\ref{eq:single-QCRB-phid}).

\section{Optimizing the BS transmission coefficients}

In the following, we start, respectively, from Eqs.~(\ref{eq:two-QCRB-phid}) and (\ref{eq:single-QCRB-phid}) to point out the optimal conditions for the BS transmission coefficient for a general family of quantum states with some parity symmetry, which requires that one of input ports is injected by an even or odd state \cite{Liu2013PRA}. This family encompasses a wide range of non-classical states employed in most current interferometric phase experiments \cite{Caves1981PRD,Holland1993PRL,Anisimov2010PRL,Luis2001PRA,Joo2011PRL,Pezze2013PRL,Tan2014PRA,Birrittella2014JOSAB,Lang2014PRA,Ouyang2016JOSAB}. If the input state is assumed to be of separable form $\vert\psi_{{\rm in}}\rangle=\vert\chi_{a}\rangle\vert\chi_{b}\rangle$ (excluding the two-mode squeezed vacuum state (TMSVS)), then Eq.~(\ref{eq:two-QCRB-phid}) can be explicitly expressed as follows:

\begin{eqnarray}
\mathcal{V}_{2}(\hat{\phi}_{d}) & \geq & \frac{1}{\upsilon}\frac{1}{\mathfrak{G}\cos^{2}\tau+\mathfrak{F}\sin^{2}\tau},\label{eq:two-QCRB-parity}
\end{eqnarray}
where
\begin{eqnarray}
\mathfrak{G} & = & \frac{4\mathcal{V}(a^{\dagger}a)\mathcal{V}(b^{\dagger}b)}{\mathcal{V}(a^{\dagger}a)+\mathcal{V}(b^{\dagger}b)},\label{eq:QCRB-G}\\
\mathfrak{F} & = & 2\langle a^{\dagger}ab^{\dagger}b\rangle+\langle a^{\dagger}a\rangle+\langle b^{\dagger}b\rangle-2{\rm Re}\left(e^{i2\vartheta}\langle a^{\dagger2}\rangle\langle b^{2}\rangle\right).\label{eq:QCRB-F}
\end{eqnarray}
Moreover, Eq.~(\ref{eq:single-QCRB-phid}) is simplified to
\begin{eqnarray}
\mathcal{V}_{1}(\hat{\phi}_{d}) & \geq & \frac{1}{\upsilon}\frac{1}{4\mathcal{V}(J_{z})\cos^{2}\tau+\mathfrak{F}\sin^{2}\tau},\label{eq:single-QCRB-parity}
\end{eqnarray}
with
\begin{eqnarray}
4\mathcal{V}(J_{z})&=&\mathcal{V}(a^{\dagger}a)+\mathcal{V}(b^{\dagger}b).
\end{eqnarray}
Here, we assume $\langle a^{2}\rangle=\left|\langle a^{2}\rangle\right|e^{i\theta_{a}}$, $\langle b^{2}\rangle=\left|\langle b^{2}\rangle\right|e^{i\theta_{b}}$ and they are nonvanishing. The detailed derivations of the above expressions are shown in Appendix A, where we also provide a more general expression for any input states without requiring a separable form. Obviously, $\mathcal{V}_{2}(\hat{\phi}_{d})=\mathcal{V}_{1}(\hat{\phi}_{d})$ for $\tau=\pi/2$ with Eqs.~(\ref{eq:two-QCRB-parity}) and (\ref{eq:single-QCRB-parity}). When $\tau\neq\pi/2$, $\mathcal{V}_{2}(\hat{\phi}_{d})\geq\mathcal{V}_{1}(\hat{\phi}_{d})$ due to
\begin{eqnarray}
\mathfrak{G}&=&\frac{4\mathcal{V}(a^{\dagger}a)\mathcal{V}(b^{\dagger}b)}{\mathcal{V}(a^{\dagger}a)+\mathcal{V}(b^{\dagger}b)}\leq\mathcal{V}(a^{\dagger}a)+\mathcal{V}(b^{\dagger}b)=4\mathcal{V}(J_{z}).
\end{eqnarray}

As a result, to maximize the sensitivities, $\mathcal{V}_{2}(\hat{\phi}_{d})$ and $\mathcal{V}_{1}(\hat{\phi}_{d})$ are equivalent to enlarge the value of the term in the denominator on the right-hand side of Eqs~(\ref{eq:two-QCRB-parity}) and (\ref{eq:single-QCRB-parity}). Let us first optimize the BS phase $\vartheta$, which is only presented in the $\mathfrak{F}$. More interestingly, according to Eq.~(\ref{eq:maxQFI}), one can see that maximizing the $\mathfrak{F}$ over $\vartheta$ is equivalent to finding the maximal variance of a component of angular momentum on the plane perpendicular to $z$ axis. From Eq.~(\ref{eq:QCRB-F}), we see that $\mathfrak{F}$ gets the maximum
\begin{eqnarray}
\mathfrak{F}&=&2\langle a^{\dagger}ab^{\dagger}b\rangle+\langle a^{\dagger}a\rangle+\langle b^{\dagger}b\rangle+2\left|\langle a^{2}\rangle\right|\left|\langle b^{2}\rangle\right|,\label{eq:QCRB-maxF}
\end{eqnarray}
when the following condition satisfies
\begin{eqnarray}
2\vartheta-\theta_{a}+\theta_{b} &=& \pm\pi.\label{eq:phase-matching}
\end{eqnarray}
This condition is quite general. It covers the specific case discussed
in Ref.~\cite{Liu2013PRA} by fixing $\vartheta=0$ (i.e., $B=\exp\left(-i\tau J_{x}\right)$).
In that work, the authors named such an optimal condition as phase-matching
condition. From Eq.~(\ref{eq:phase-matching}), one can see that
when the expansion coefficients of input state are real, $\vartheta=\pm\pi/2$
(i.e., $B=\exp\left(-i\tau J_{y}\right)$) is optimal as considered
in Ref.~\cite{Zhong2017PRA}. Note that an arbitrary value of phase
for the BS is optimal if $\langle a^{2}\rangle$ or $\langle b^{2}\rangle$
vanishes.

Apart from the phase $\vartheta$, a brief glance at Eqs.~(\ref{eq:two-QCRB-parity}) and (\ref{eq:single-QCRB-parity}) suffices to find that $\mathcal{V}_{2}(\hat{\phi}_{d})$ and $\mathcal{V}_{1}(\hat{\phi}_{d})$ can be further enhanced with optimization of $\tau$, equivalently, the transmission ratio of the BS. To gain the minimum $\mathcal{V}_{2}(\hat{\phi}_{d})$, the optimal $\tau$ accounts for a comparison between $\mathfrak{G}$ and $\mathfrak{F}$ from Eq.~(\ref{eq:two-QCRB-parity}). Similarly, the optimal $\tau$ for $\mathcal{V}_{1}(\hat{\phi}_{d})$ accounts for a comparison between $4\mathcal{V}(J_{z})$ and $\mathfrak{F}$ from Eq.~(\ref{eq:single-QCRB-parity}).

To understand more the optimization of $\tau$ and $\vartheta$,
below, we consider various specific states that have been frequently
investigated in quantum-enhanced interferometry.

\emph{Case 1.}\textemdash Let us first consider two special cases, where the input states are chosen as a combination of two sub-Poissonian states (for instance, twin Fock state (TFS) $\vert\kappa\rangle\vert\kappa\rangle$, which is known as the Holland-Burnett state for $\tau=\pi/2$ \cite{Holland1993PRL}) and a mixing of CS $\vert\alpha\rangle$ with a sub-Poissonian state (for instance, $\vert\alpha\rangle\vert\kappa\rangle$ \cite{Pezze2013PRL}). For these cases, $\tau_{{\rm opt}}=\pi/2$ (i.e., a $50:50$ BS) is optimal since
\begin{eqnarray}
\mathfrak{G}&\leq&4\mathcal{V}(J_{z})\leq n_{a}+n_{b}<\mathfrak{F}.
\end{eqnarray}
This can also be straightly derived. As for Fock states, the variance of photon number operator is vanishing, and it then yields, according to Eq.~(\ref{eq:QCRB-G}), $\mathfrak{G}=0$. This implies that the sensitivity of Eq.~(\ref{eq:two-QCRB-parity}) is only determined by Eq.~(\ref{eq:QCRB-F}). In Eq.~(\ref{eq:QCRB-F}), one can further observe that the expectation value of the operator $b^2$ in Fock states is vanishing. This means that the result of $\mathfrak{F}$ of Eq.~(\ref{eq:QCRB-F}) is independent on the phase of the BS. Thus, a balanced BS of arbitrary phase is optimal for both TFS and CS$\otimes$FS cases. See Table~\ref{tab:QFI} for the ultimate sensitivities given by $\vert\kappa\rangle\vert\kappa\rangle$ and $\vert\alpha\rangle\vert\kappa\rangle$ with $n_{i}=\kappa$ for $\vert\kappa\rangle_{i},\thinspace\left(i=a,b\right)$ and $n_{a}=\left|\alpha\right|^{2}$ for $\vert\alpha\rangle_{a}$.

\emph{Case 2.}\textemdash We then consider the input state to be a mixing of a CS $\vert\alpha\rangle$ with a coherent superposition state (CSS)
\begin{eqnarray}
\vert{\rm CSS}\rangle&=&\left(\vert\beta\rangle+\vert-\beta\rangle\right)/\sqrt{\mathcal{N}_{{\rm CSS}}}
\end{eqnarray}
with $\beta=\left|\beta\right|e^{i\varphi_{\beta}}$ and the normalization factor $\mathcal{N}_{{\rm CSS}}=2(1+e^{-2\left|\beta\right|^{2}})$. It has been demonstrated that it brings more advantage to sensitivity for single-arm phase measurement with the case of $\alpha=\beta$ \cite{Joo2011PRL}. Such advantage stems from a high degree of mode entanglement of the resultant state created by the interference between $\vert\alpha\rangle$ and $\vert{\rm CSS}\rangle$, which is recognized as a superposition of NOON state.

For a CS $\vert\alpha\rangle$, one has $\mathcal{V}(a^{\dagger}a)=n_{a}=\left|\alpha\right|^{2}$ and $\langle a^{2}\rangle=\alpha^{2}$. For a CSS, one gets $\langle b^{2}\rangle=\beta^{2}$, and the mean of photon number $\langle b^{\dagger}b\rangle$ equals the one of CS up to a factor $\tanh(\left|\beta\right|^{2})$. When $\beta\gg1$ (namely, $\tanh(\left|\beta\right|^{2})\rightarrow1$), we then have $\mathcal{V}(b^{\dagger}b)=n_{b}=\left|\beta\right|^{2}$, asymptotically. Submitting these exact expressions into Eqs.~(\ref{eq:QCRB-G}) and (\ref{eq:QCRB-maxF}) yields the ultimate phase sensitivity defined by Eq.~(\ref{eq:two-QCRB-parity}) (see Table~\ref{tab:QFI}). Obviously, a balanced BS ($\tau_{{\rm opt}}=\pi/2$) is optimal for the CS$\otimes$ CSS due to
\begin{eqnarray}
\mathfrak{G}&\leq&4\mathcal{V}(J_{z})=n_{a}+n_{b}<\mathfrak{F},
\end{eqnarray}
and the optimal BS phase reads $\vartheta_{{\rm opt}}=(2\varphi_{\alpha}-2\varphi_{\beta}\pm\pi)/2$.

\emph{Case 3.}\textemdash As one of the most celebrated strategies for sub-shot-noise interferometry, the use of $\vert\alpha\rangle\vert\xi\rangle$ was proposed by Caves nearly 40 years ago \cite{Caves1981PRD}. Nowadays, it has been implemented in the development of next-generation gravitational-wave detector \cite{Collaboration2011NP,Collaboration2013NP}.

As for SVS $\vert\xi\rangle$ with $\xi=\left|\xi\right|e^{i\varphi_{\xi}}$, the variance of photon number is $\mathcal{V}(b^{\dagger}b)=\frac{1}{2}\sinh^{2}2\left|\xi\right|$, and the mean of the squared annihilation operator is $\langle b^{2}\rangle=-\frac{1}{2}\sinh2\left|\xi\right|e^{i\varphi_{\xi}}$. Rewriting these expressions in terms of $n_{b}$ under the consideration of $n_{b}=\sinh^{2}\left|\xi\right|$ as $\mathcal{V}(b^{\dagger}b)=2n_{b}\left(1+n_{b}\right)$ and $\langle b^{2}\rangle=\sqrt{n_{b}\left(1+n_{b}\right)}$, then submitting them into Eqs.~(\ref{eq:QCRB-G}) and (\ref{eq:QCRB-maxF}) finally gives the ultimate phase sensitivities of Eqs.~(\ref{eq:two-QCRB-parity}) and (\ref{eq:single-QCRB-parity})(see Table~\ref{tab:QFI}) \cite{Jarzyna2012PRA,Liu2013PRA}. One can check that $\mathfrak{G}$ is still less than $\mathfrak{F}$ for an arbitrary mean number of photons so that a balanced BS is again the best choice for the case of CS$\otimes$SVS. Correspondingly, the optimal BS phase for $\mathcal{V}_{2}(\hat{\phi}_{d})$ reads $\vartheta_{{\rm opt}}=(2\varphi_{\alpha}-\varphi_{\xi})/2$. As for $\mathcal{V}_{1}(\hat{\phi}_{d})$, the situation is complicated by the fact that the inequality $4\mathcal{V}(J_{z})<\mathfrak{F}$ holds only under the condition of $n_{b}<2n_{a}$, while $4\mathcal{V}(J_{z})>\mathfrak{F}$ when $n_{b}\geq2n_{a}$. Thus, the maximal sensitivity of $\mathcal{V}_{1}(\hat{\phi}_{d})$ is given by
\begin{eqnarray}
\mathcal{V}_{1}(\hat{\phi}_{d})&=&\begin{cases}
(\upsilon\mathfrak{F})^{-1}, & n_{b}<2n_{a},\\{}
[4\upsilon\mathcal{V}(J_{z})]^{-1}, & n_{b}\geq2n_{a}.
\end{cases}
\end{eqnarray}
Here, $\mathcal{V}_{1}(\hat{\phi}_{d})$ gets the minimum at $\vartheta_{{\rm opt}}=0$ and $\tau_{{\rm opt}}=(2\varphi_{\alpha}-\varphi_{\xi})/2$ with $n_{b}\leq2n_{a}$; however, when $n_{b}\geq2n_{a}$, the optimal BS transmission ratio and phase are $\tau_{{\rm opt}}=0$ and $\vartheta_{{\rm opt}}\in[-\pi,\pi]$ (see Table~\ref{tab:QFI}). It is remarkable that the two-mode optical interferometer with $\tau=0$ is equivalent to two independent single-mode phase estimation \cite{Monras2006PRA,Genoni2011PRL}.

Now, we consider a modified strategy of Caves by powering the interferometer with two identical SVSs $\vert\xi\rangle\vert\xi\rangle$; we call it as twin squeezed vacuum state. Interestingly, in such case, we find $\mathfrak{G}=4\mathcal{V}(J_{z})=\mathfrak{F}$. This implies that for both $\mathcal{V}_{2}(\hat{\phi}_{d})$ and $\mathcal{V}_{1}(\hat{\phi}_{d})$, the BS with arbitrary transmission ratio is optimal. For a more general nonidentical case $\vert\xi\rangle\vert\xi^{\prime}\rangle$ with
$\xi\neq\xi^{\prime}$ \cite{Lang2014PRA} and we obtain
\begin{eqnarray}
\mathfrak{G} & = & \frac{8n_{a}\left(1+n_{a}\right)n_{b}\left(1+n_{b}\right)}{n_{a}\left(1+n_{a}\right)+n_{b}\left(1+n_{b}\right)},\\
4\mathcal{V}(J_{z}) & = & 2n_{a}\left(1+n_{a}\right)+2n_{b}\left(1+n_{b}\right),\\
\mathfrak{F} & = & 2n_{a}n_{b}+n_{a}+n_{b}+2\sqrt{n_{a}\left(1+n_{a}\right)}\sqrt{n_{b}\left(1+n_{b}\right)}.
\end{eqnarray}
 With these, we find that $\mathcal{V}_{2}(\hat{\phi}_{d})$ reaches the minimum
\begin{eqnarray}
\mathcal{V}_{2}(\hat{\phi}_{d})&=&(\upsilon\mathfrak{F})^{-1}
\end{eqnarray}
at $\tau_{{\rm opt}}=\pi/2$ and $\vartheta_{{\rm opt}}=(\varphi_{\xi}-\varphi_{\xi^{\prime}}\pm\pi)/2$, due to $\mathfrak{G}<\mathfrak{F}$. While for $\mathcal{V}_{1}(\hat{\phi}_{d})$, $\tau=\pi/2$ is the worst option due to $\mathfrak{F}<4\mathcal{V}(J_{z})$. Hence, the maximal sensitivity of $\mathcal{V}_{1}(\hat{\phi}_{d})$
is given by
\begin{eqnarray}
\mathcal{V}_{1}(\hat{\phi}_{d})&=&[4\upsilon\mathcal{V}(J_{z})]^{-1},
\end{eqnarray}
under the optimal conditions of $\tau_{{\rm opt}}=0$ and $\vartheta_{{\rm opt}}\in[-\pi,\pi]$.

\emph{Case 4.}\textemdash Finally, we consider the TMSVS,
\begin{eqnarray}
\left|\psi_{{\rm in}}\right\rangle  & = & \exp\left(\zeta^{\ast}ab-\zeta a^{\dagger}b^{\dagger}\right)\vert00\rangle,
\end{eqnarray}
which can be understood as a superposition of TFSs equipped with $n_{a}=n_{b}=\sinh^{2}\left|\zeta\right|$
\cite{Gerry2004Book}. Although the TMSVS cannot be written as a separable form like those given above and thus Eq.~(\ref{eq:two-QCRB-parity}) does not hold here, it also satisfies the property of parity symmetry with $\left\langle a\right\rangle =\left\langle b\right\rangle =0$. According to Eqs.~(\ref{eq:FsdM0}) and (\ref{eq:FddM0}), one can easily find that $F_{sd}=0$ and $F_{dd}=\mathfrak{F}\sin^{2}\tau$ and hence
\begin{eqnarray}
\mathcal{V}_{2}(\hat{\phi}_{d})&=&\mathcal{V}_{1}(\hat{\phi}_{d})=(\upsilon\mathfrak{F}\sin^{2}\tau)^{-1},
\end{eqnarray}
according to Eqs.~(\ref{eq:two-QCRB-phid}) and (\ref{eq:single-QCRB-phid}). Assuming the total mean number of photons as $n_{{\rm tot}}=n_{a}+n_{b}$, the sensitivities $\mathcal{V}_{2}(\hat{\phi}_{d})$ and $\mathcal{V}_{1}(\hat{\phi}_{d})$ are maximized as $(n_{{\rm tot}}^{2}+2n_{{\rm tot}})^{-1}$ by a $50:50$ BS (see Table~\ref{tab:QFI}) \cite{Anisimov2010PRL}. Interestingly, it is irrelevant to the phase of the BS, namely, an arbitrary value of $\vartheta$ for the BS is optimal.

\begin{figure}[H]
\centering
\includegraphics[width=1\columnwidth]{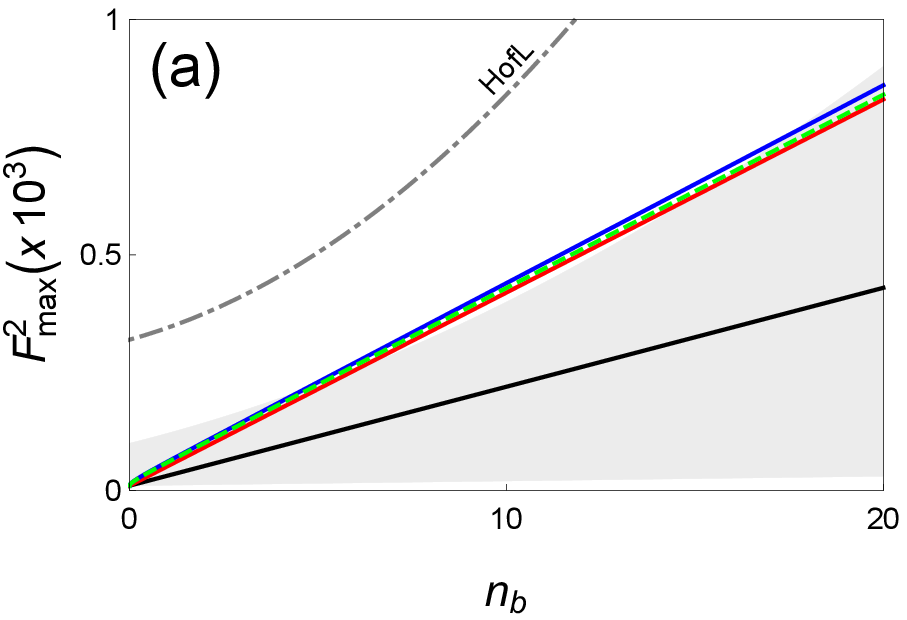}
\includegraphics[width=1\columnwidth]{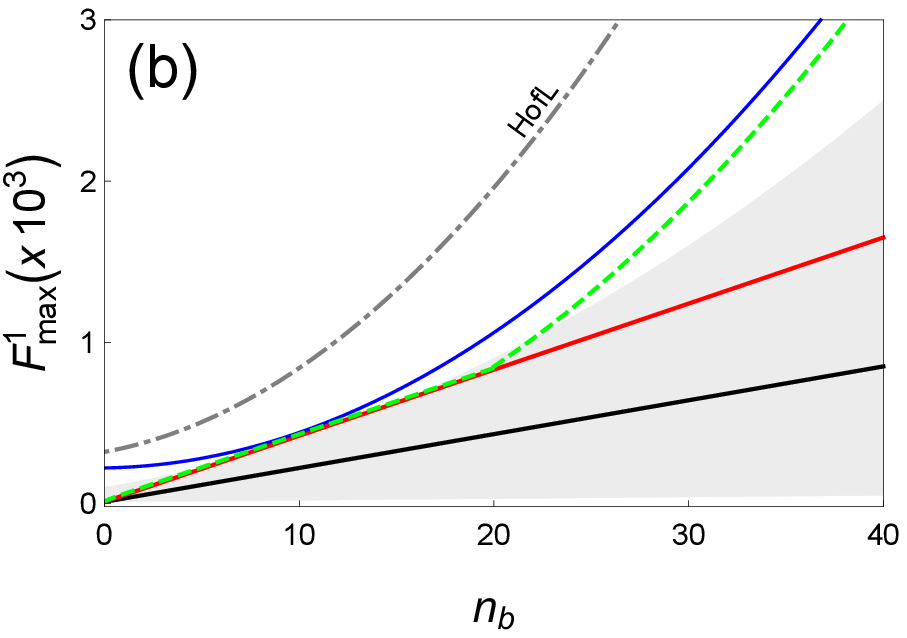}
\caption{(Color online) The maximal effective QFIs $F_{{\rm max}}^{i}\equiv1/[\upsilon\mathcal{V}_{i}(\hat{\phi}_{d})]$, $(i=1,2)$ as a function of $n_{b}$ for different non-classical states listed in Table~\ref{tab:QFI} with fixed $n_{a}=10$. (a) is for $i=2$, $F_{{\rm max}}^{2}=\mathfrak{F}$ and (b) for $i=1$, $F_{{\rm max}}^{1}=\max[4\mathcal{V}(J_{z}),\mathfrak{F}]$. The black-solid, red, green-dashed, and blue-solid lines represent CS$\otimes$FS $\vert\alpha\rangle\vert\kappa\rangle$, CS$\otimes$CSS
$\vert\alpha\rangle\vert{\rm CSS}\rangle$, CS$\otimes$SVS $\vert\alpha\rangle\vert\xi\rangle$, and two SVSs $\vert\xi\rangle\vert\xi^{\prime}\rangle$, respectively. The gray dot-dashed line corresponds to the Hofmann limit for two SVSs \cite{Hofmann2009PRA}. The shaded area represents the sub-shot-noise sensitivity region bounded by $1/\langle N\rangle$ and $1/\langle N\rangle^{2} $.}
\label{fig:CS}
\end{figure}

According to Table~\ref{tab:QFI}, a hierarchy for performance of the above non-classical states in phase sensitivity is given as follows:
\begin{equation}
{\rm TFS}\!=\!{\rm CS}\otimes{\rm FS}\!<\!{\rm CS}\otimes{\rm CSS}\!<\!{\rm CS}\otimes{\rm SVS}\!<\!{\rm TSVS}\!=\!{\rm TMSVS},\label{eq:hierarchy}
\end{equation}
by restricting the mean photon numbers on each arm to be equal. There are two pairs of equivalence, TFS and CS$\otimes$FS, as well as TSVS and TMSVS. We see that both the TSVS and the TMSVS give the best performance for sensitivity.

We plot in Fig.~\ref{fig:CS} the maximal effective QFIs $F_{{\rm max}}^{i}\equiv1/[\upsilon\mathcal{V}_{i}(\hat{\phi}_{d})]$, $(i=1,2)$ as a function of \textbf{$n_{b}$}{} with a fixed $n_{a}=10$. As shown in Fig.~\ref{fig:CS}(a), the sensitivities given by CS$\otimes$CSS, CS$\otimes$SVS, and two SVSs almost emerged. Excitingly, except TFS and CS$\otimes$FS, all other states beat the bound of $1/\left\langle N\right\rangle^2 $ in the interval near the point $n_{b}=n_{a}$. A similar phenomenon is also observed for $\mathcal{V}_{1}(\hat{\phi}_{d})$ shown in Fig.~\ref{fig:CS}(b). Besides near the point $n_{b}=n_{a}$, we see that two SVSs beat $1/\left\langle N\right\rangle^2 $ for any $n_{b}$, and CS$\otimes$SVS does so when $n_{b}>2n_{a}$. This seems to contradict the fact that the HL is the fundamental sensitivity limit for linear optical interferometry. We note that such counterintuitive behavior is caused by the problematic definition of the HL. It is true for the fundamental limit for states with a fixed photon number, but it is false for the cases with a fluctuating photon number. To conquer this problem, Hofmann suggested a variant form of HL, which is defined in terms of averaging over the squared photon number $1/\langle N^{2}\rangle$ \cite{Hofmann2009PRA}, which was further discussed in \cite{Hyllus2010PRL,Pezze2015PRA}. It is clearly shown in Fig.~\ref{fig:CS} that such limit cannot be beaten.

Up to now, we have not discussed how to attain the above ultimate sensitivities with some specific feasible measurements. It was shown that when the probe state\textemdash the state prior to the phase shift operation\textemdash is symmetric pure state, double-output-port photon number counting measurement \cite{Hofmann2009PRA,Krischek2011PRL,Lang2013PRL,Zhong2017PRA} is globally optimal over the whole range of phase interval. Fortunately, all parity symmetric states here belong to such a case when $\tau=\pi/2$. While for $\tau=0$, the optimal measurements were addressed in Refs.~\cite{Monras2006PRA,Genoni2011PRL}. Therefore, the ultimate sensitivities for $\mathcal{V}_{2}(\hat{\phi}_{d})$ and $\mathcal{V}_{1}(\hat{\phi}_{d})$ listed in Table~\ref{tab:QFI} are always saturated.

\section{Sensitivities enhanced by photon addition and photon subtraction}

Photon addition and subtraction have attracted much attention since the resulting state generated by adding or subtracting photons to a state may create a higher degree of entanglement than the initial state \cite{Kitagawa2006PRA,Dakna1997PRA,Kim2008JPB,Wang2013EPJD,Navarrete-Benlloch2012PRA,Dakna1999PRA,Fiurasek2005PRA,Biswas2007PRA,Ourjoumtsev2007PRL}. This attractive property inspires substantial interest to implement such novel states into quantum information processing \cite{Duan2000PRL,Eisert2004AP}, quantum key distribution \cite{Opatrny2000PRA,Cochrane2002PRA,Olivares2003PRA,Wenger2004PRL}, and even quantum metrology \cite{Braun2014PRA,Birrittella2014JOSAB,Tan2014PRA,Wang2019OC}. Photon addition or subtraction can be simply achieved by a weak interaction with an ancillary mode. The detection of a photon in this additional mode indicates a successful addition or subtraction event \cite{Zavatta2004Sci,Ourjoumtsev2007PRL}. The interaction process of adding photons is applied by a weakly reflecting BS, and that of subtracting photons is implemented by a weak parametric amplifier. Then, a multiple photon addition or subtraction can be realized by a number of single-photon addition or subtraction events occurring. As the number of addition or subtraction events increases, the probability for successfully adding or subtracting the photons decreases rapidly.

In a recent work \cite{Lang2013PRL}, Lang and Caves observed that when one of the input ports of the interferometer with a balanced BS $(\tau=\pi/2)$ is injected by a CS, the SVS is the best choice to put to the interferometer's secondary input port. Recently, Birrittella and Gerry subsequently demonstrated that the phase sensitivity given by CS$\otimes$PSSVS may outperform the one given by CS$\otimes$SVS under the same values of the CS amplitude and the squeezing parameter \cite{Birrittella2014JOSAB}. More recently, Wang \emph{et al. }found that when the phase shift to be estimated approaches zero, the CS$\otimes$SVS is indeed the optimal state under a constraint on the average photon number. However, when the phase shift slightly deviates from zero, in terms of parity detection, CS$\otimes$PASVS can give the better phase sensitivity than both CS$\otimes$SVS and CS$\otimes$PSSVS \cite{Wang2019OC}. It remains unclear whether CS$\otimes$PASVS and CS$\otimes$PSSVS may outperform CS$\otimes$SVS for the phase sensitivity. This allows us to revisit the scenarios considered in \cite{Birrittella2014JOSAB,Wang2019OC}. To clarify this issue, we need to calculate the maximal QFI for CS$\otimes$PASVS and CS$\otimes$PSSVS.

Formally adding and subtracting $\kappa$ photons to a SVS $\vert\xi\rangle$ can be represented, respectively, as
\begin{eqnarray}
\vert\xi,\kappa^{+}\rangle=\frac{b^{\dagger\kappa}\vert\xi\rangle}{\sqrt{\mathcal{N}_{\xi,\kappa^{+}}}}, & \quad & \vert\xi,\kappa^{-}\rangle=\frac{b^{\kappa}\vert\xi\rangle}{\sqrt{\mathcal{N}_{\xi,\kappa^{-}}}}.\label{eq:pm-SVS}
\end{eqnarray}
Here and below, we use the symbols $+$ and $-$ to indicate the photon addition and subtraction, respectively. Obviously, states of Eq.~(\ref{eq:pm-SVS}) satisfy parity symmetry. For simplicity, we assume the squeezing parameter $\xi$ to be a real constant. The normalization coefficients for PASVS and PSSVS are explicitly obtained by
\begin{eqnarray}
\mathcal{N}_{\xi,\kappa^{\pm}} & = & x_{\pm}^{\kappa}\sum_{l=0}^{\left[\frac{\kappa}{2}\right]}\frac{\kappa!\kappa!}{l!l!\left(\kappa-2l\right)!}\left(\frac{y}{x_{\pm}}\right)^{2l},\label{eq:normalization-1}
\end{eqnarray}
associating with
\begin{eqnarray}
x_{+}\equiv\cosh^{2}\xi\text{,} & \thinspace\thinspace x_{-}\equiv\sinh^{2}\xi, & \thinspace\thinspace{\rm and}\thinspace\thinspace y\equiv\frac{1}{4}\sinh2\xi.
\end{eqnarray}
With the help of Legendre polynomials, Eq.~(\ref{eq:normalization-1}) can be rewritten in a more concise form as
\begin{eqnarray}
\mathcal{N}_{\xi,\kappa^{+}}&=&\kappa!\left(\cosh^{\kappa}\xi\right)P_{\kappa}\!\left(\cosh\xi\right),
\end{eqnarray}
\cite{Hong-Yi2006CTP,Hu2010JMO} and
\begin{eqnarray}
\mathcal{N}_{\xi,\kappa^{-}}&=&\kappa!\left[-\left(i\sinh\xi\right)\right]^{\kappa}P_{\kappa}\left(i\sinh\xi\right),
\end{eqnarray}
\cite{Meng2012JOS,HONG-YI2019MPLA}. With Eq.~(\ref{eq:normalization-1}), the mean photon number operator in terms of these states is given by
\begin{eqnarray}
n_{b}^{+}=\frac{\mathcal{N}_{\xi,\left(\kappa+1\right)^{+}}-\mathcal{N}_{\xi,\kappa^{+}}}{\mathcal{N}_{\xi,\kappa^{+}}}, & \quad & n_{b}^{-}=\frac{\mathcal{N}_{\xi,\left(\kappa+1\right)^{-}}}{\mathcal{N}_{\xi,\kappa^{-}}}.\label{eq:PS-1-1}
\end{eqnarray}
To obtain the explicit expressions of Eqs.~(\ref{eq:two-QCRB-parity}) and (\ref{eq:single-QCRB-parity}) for $\vert\alpha\rangle\vert\xi,\kappa^{+}\rangle$ and $\vert\alpha\rangle\vert\xi,\kappa^{-}\rangle$, we also need to calculate the average values of the squared photon number
\begin{eqnarray}
\langle(b^{\dagger}b)^{2}\rangle_{+} & = & \frac{\mathcal{N}_{\xi,\left(\kappa+2\right)^{+}}-3\mathcal{N}_{\xi,\left(\kappa+1\right)^{+}}+\mathcal{N}_{\xi,\kappa^{+}}}{\mathcal{N}_{\xi,\kappa^{+}}},\label{eq:PS-2}\\
\langle(b^{\dagger}b)^{2}\rangle_{-} & = & \frac{\mathcal{N}_{\xi,\left(\kappa+2\right)^{-}}+\mathcal{N}_{\xi,\left(\kappa+1\right)^{-}}}{\mathcal{N}_{\xi,\kappa^{-}}},\label{eq:PS-3}
\end{eqnarray}
and the mean squared annihilation operator

\begin{eqnarray}
\left\langle b^{2}\right\rangle _{\pm}&=&\frac{x_{\pm}^{\kappa+1}}{\mathcal{N}_{\xi,\kappa^{\pm}}}\sum_{l=0}^{\left[\frac{\kappa}{2}\right]}\frac{\kappa!\left(\kappa+2\right)!}{l!\left(l+1\right)!\left(\kappa-2l\right)!}\left(-\frac{y}{x_{\pm}}\right)^{2l+1}.\label{eq:PS-4}
\end{eqnarray}
Submitting above Eqs.~(\ref{eq:PS-1-1}), (\ref{eq:PS-2}), (\ref{eq:PS-3}), and (\ref{eq:PS-4}) to Eqs.~(\ref{eq:two-QCRB-parity}) and (\ref{eq:single-QCRB-parity}) finally yields the ultimate phase sensitivities given by $\vert\alpha\rangle\vert\xi,\kappa^{+}\rangle$ and $\vert\alpha\rangle\vert\xi,\kappa^{-}\rangle$.

\begin{figure}[H]
\centering
\includegraphics[width=1\columnwidth]{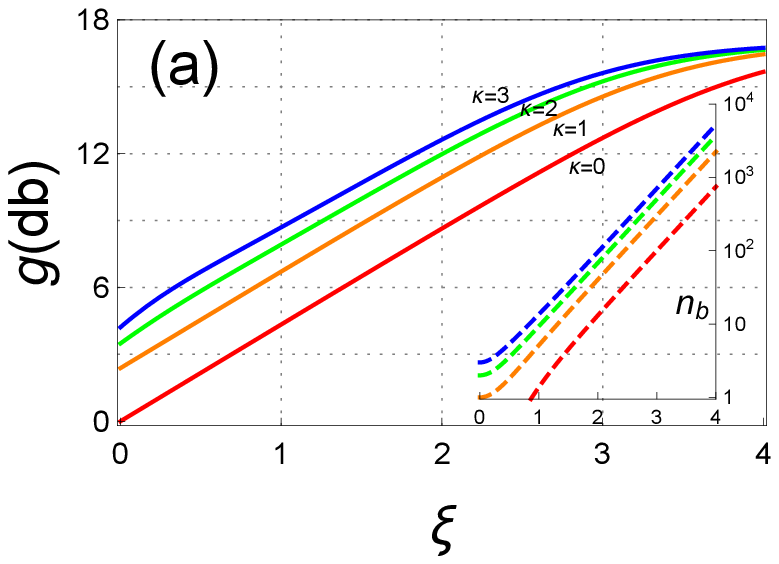}
\includegraphics[width=1\columnwidth]{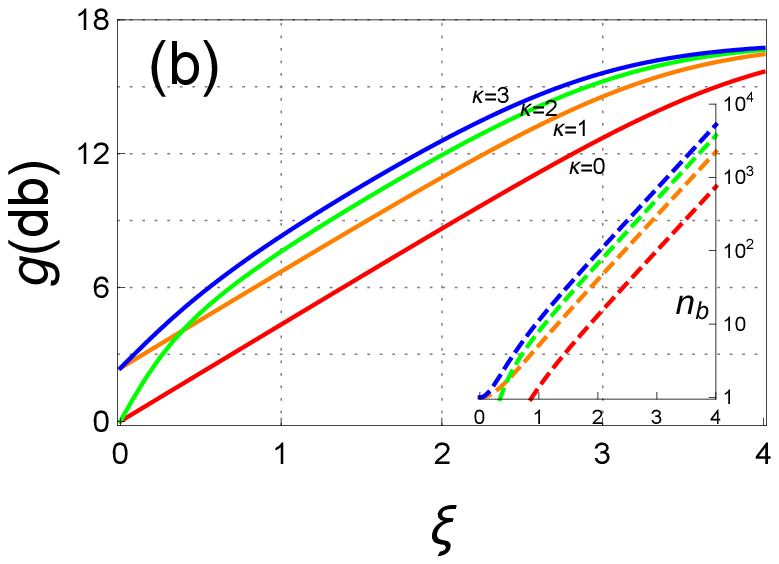}
\caption{(Color online) Phase sensitivity gain $g\equiv-10\log_{10}(\mathcal{V}(\hat{\phi}_{d})\sqrt{\upsilon\left\langle N\right\rangle })$ as a function of the squeezing parameter $\xi$ for CS$\otimes$PASVS $\vert\alpha\rangle\vert\xi,\kappa^{+}\rangle$ (a) and CS$\otimes$PSSVS $\vert\alpha\rangle\vert\xi,\kappa^{-}\rangle$ (b). Different color curves correspond to different photon number $\kappa$ added or subtracted to the SVS. Insert corresponds to the average photon number $n_{b}$ for PASVS or PSSVS vs $\xi$.}
\label{fig:CS-1}
\end{figure}

We plot in Fig.~\ref{fig:CS-1} the phase sensitivity gain
\begin{eqnarray}
g&\equiv&-10\log_{10}\left(\mathcal{V}(\hat{\phi}_{d})\sqrt{\upsilon\left\langle N\right\rangle }\right),
\end{eqnarray}
for $\tau=\pi/2$ as a function of the squeezing parameter $\xi$ with $\alpha=25$ for different added (subtracted) photon numbers from $0$ to $3$. Clearly, $\kappa=0$ indicates the CS$\otimes$SVS $\vert\alpha\rangle\vert\xi\rangle$ \cite{Caves1981PRD,Pezze2008PRL,Jarzyna2012PRA}. For both cases of $\vert\alpha\rangle\vert\xi,\kappa^{+}\rangle$ and $\vert\alpha\rangle\vert\xi,\kappa^{-}\rangle$, the level of sensitivity gain increases significantly from $0$ to $1$ and gradually decreases as $\kappa$ increases. This phenomenon is strongly related to the great increase of photon numbers in $\vert\xi,\kappa^{+}\rangle$ and $\vert\xi,\kappa^{-}\rangle$ \cite{Birrittella2014JOSAB}, as shown in insert of Fig.~\ref{fig:CS-1}(a) and (b). The sensitivity gains with different $\kappa$ go asymptotically the same level when $\xi$ increases to a higher degree such that $n_{b}^{\pm}\gg n_{a}$. We observe a different phenomenon for the cases of photon addition and photon subtraction with the squeezing parameter $\xi$ at a low range of $0\sim1$. It is worth to note that the sensitivity for $\vert\alpha\rangle\vert\xi,\kappa^{+}\rangle$ equals to that for $\vert\alpha\rangle\vert\kappa\rangle$ (see Table~\ref{tab:QFI}) at the point of $\xi=0$, in which the photon-added SVS $\vert\xi,\kappa^{+}\rangle$ reduces to a Fock state $\vert\kappa\rangle$. Taking a care glance at Fig.~\ref{fig:CS-1}(a) and (b), we see that both $\vert\xi,\kappa^{+}\rangle$ and $\vert\xi,\kappa^{-}\rangle$ seem to provide the same level of sensitivity gain when $\xi>1$, while for $\kappa=1$ and $2$, they are the same for arbitrary value of $\xi$.

To evaluate the effect of CS$\otimes$PASVS and CS$\otimes$PSSVS on the phase sensitivity for a given mean photon number of the input state, we take a typical case of $\kappa=1$, for example. From Eq.~(\ref{eq:pm-SVS}), one-single-photon-added and one-single-photon-subtracted SVSs are exactly represented, respectively, as follows:

\begin{eqnarray}
\vert\xi,1^{+}\rangle=\frac{b^{\dagger}\vert\xi\rangle}{\cosh\xi}, & \quad & \vert\xi,1^{-}\rangle=\frac{b\vert\xi\rangle}{\sinh\xi}.
\end{eqnarray}
The mean photon numbers $n_{b}$ of these states are given, respectively, by

\begin{eqnarray}
n_{b}^{+}=3\cosh^{2}\xi-2, & \quad & n_{b}^{-}=3\sinh^{2}\xi+1.\label{eq:PS1-1}
\end{eqnarray}
Having these equations and combining Eqs.~(\ref{eq:PS-2}), (\ref{eq:PS-3}), and (\ref{eq:PS-4}), we get the exact solutions of $\mathcal{V}(b^{\dagger}b)$ and $\langle b^{2}\rangle$ in terms of $n_{b}^{+}$ and $n_{b}^{-}$. Interestingly, both $\vert\xi,1^{+}\rangle$ and $\vert\xi,1^{-}\rangle$ give the same expressions as $\mathcal{V}(b^{\dagger}b)=\frac{2}{3}\left(n_{b}^{2}+n_{b}-2\right)$ and $\langle b^{2}\rangle=-\sqrt{n_{b}^{2}+n_{b}-2}$. Note that here, we neglect the symbols $\pm$ in $n_{b}^{\pm}$ for simplicity. Hence, with Eq.~(\ref{eq:two-QCRB-parity}), the maximal sensitivity of $\mathcal{V}_{2}(\hat{\phi}_{d})$ is given by
\begin{eqnarray}
\mathcal{V}_{2}(\hat{\phi}_{d})&=&\left(\upsilon\mathfrak{F}\right)^{-1},
\end{eqnarray}
associating with
\begin{eqnarray}
\mathfrak{F} & = & 2n_{a}n_{b}+n_{a}+n_{b}+2n_{a}\sqrt{n_{b}^{2}+n_{b}-2},\label{eq:QCRB-F-SVS1}
\end{eqnarray}
by choosing $\vartheta_{{\rm opt}}=0$ according to Eq.~(\ref{eq:phase-matching}) and $\tau_{{\rm opt}}=\pi/2$ due to
\begin{eqnarray}
\mathfrak{G} & = & \frac{8}{3}\frac{n_{a}\left(n_{b}^{2}+n_{b}-2\right)}{n_{a}+\frac{2}{3}\left(n_{b}^{2}+n_{b}-2\right)}<\mathfrak{F}.\label{eq:QCRB-G-SVS1}
\end{eqnarray}
Obviously, Eq.~(\ref{eq:QCRB-F-SVS1}) is slightly less than that given by CS$\otimes$SVS shown in Table~\ref{tab:QFI}. The same result also holds for $\mathcal{V}_{1}(\hat{\phi}_{d})$ given in Eq.~(\ref{eq:single-QCRB-parity}), as the maximal sensitivity of $\mathcal{V}_{1}(\hat{\phi}_{d})$ is given by
\begin{eqnarray}
\mathcal{V}_{1}(\hat{\phi}_{d})&=&\begin{cases}
(\upsilon\mathfrak{F})^{-1}, & n_{b}\leq6n_{a},\\{}
[4\upsilon\mathcal{V}(J_{z})]^{-1}, & n_{b}>6n_{a}.
\end{cases}\label{eq:single-sensitivity-ASVS}
\end{eqnarray}
with
\begin{eqnarray}
4\mathcal{V}(J_{z})&=&n_{a}+\frac{2}{3}\left(n_{b}^{2}+n_{b}-2\right).
\end{eqnarray}
Obviously, the above expression of $4\mathcal{V}(J_{z})$ is also less than that of CS$\otimes$SVS shown in Table~\ref{tab:QFI}. Similar to the case of CS$\otimes$SVS, the $\mathcal{V}_{1}(\hat{\phi}_{d})$ for $1$-PASVS or $1$-PSSVS reaches the minimum at $\vartheta_{{\rm opt}}=0$ and $\tau_{{\rm opt}}=\pi/2$ when $n_{b}\leq6n_{a}$. When $n_{b}>6n_{a}$, $\tau=\pi/2$ is the worst choice, the optimal conditions are $\tau_{{\rm opt}}=0$ and $\vartheta_{{\rm opt}}\in[-\pi,\pi]$. All these results indicate that the mixing of a CS with $1$-PASVS or $1$-PSSVS does not give a higher sensitivity than CS$\otimes$SVS under the constraint of the mean total photon number.

Besides, we note that producing photon-added or photon-subtracted states is necessarily a probabilistic process with, typically, a low probability of success \cite{Barnett2018PRA}. This low success probability of state production may further increase the number of photons consumed in realistic experiments. Thus, if considering the total source consumed in experiments, the use of CS$\otimes$PASVS and CS$\otimes$PSSVS may be too expensive for high-precision phase measurement.

The work done by Wang \emph{et al.} \cite{Wang2019OC} showed that the conclusion of Lang and Caves holds only in the asymptotic limit $\phi_{d}\rightarrow0$. When the phase shift slightly departs from the zero point, they found that the PASVS outperforms both the SVS and the PSSVS. We note such a contradictory finding step from the specific detection method, i.e., parity measurement, and they have chosen \cite{Wang2019OC}. It was shown that parity detection is not a global optimal measurement, which can rarely saturate the QCRB at particular values of the phase shift \cite{Seshadreesan2013PRA}. As discussed at the end of Sec. III, for all symmetry pure states, two-output-port photon number measurement can access the full interval value of the phase shift when $\tau=\pi/2$ \cite{Hofmann2009PRA,Krischek2011PRL,Lang2013PRL,Zhong2017PRA}. Since both $\vert\alpha\rangle\vert\xi,\kappa^{+}\rangle$ and $\vert\alpha\rangle\vert\xi,\kappa^{-}\rangle$ satisfy this condition, thus the sensitivities for them can always be saturated with this two-output-port measurement. Therefore, if one uses a two-output-port photon number measurement, CS$\otimes$SVS always outperforms CS$\otimes$PASVS and CS$\otimes$PSSVS in the whole range of $\phi_{d}$.

\section{Conclusion}

We have analyzed the phase sensitivities in MZI with asymmetric BSs. Based on the single- and multi-parameter estimation theory, respectively, we analytically derive the ultimate sensitivities for a broad family of quantum input states with parity symmetry. We obtain the optimal conditions for the transmission ratio and the phase of the BS to obtain the maximal phase accuracy. We also apply these conditions to variousnon-classical states. Based on the multi-parameter estimation theory, the highest sensitivities are obtained with a balanced BS for most of these states, and the optimal BS phase depends conditionally on the specific type of probe state. According to single-parameter estimation theory, the things become more complicated, and the optimal conditions for certain states are inconsistent to that according to the multi-parameter estimation theory. Taking the CS$\otimes$SVS, for instance, a balanced BS is optimal for $n_{b}<2n_{a}$, while it is the worst option for $n_{b}\geq2n_{a}$, when a complete transmission of the BS is the best. Interestingly, both two estimation theories suggest that the sensitivity given by the TSVS is irrelevant to the transmission ratio of the BS, and the BS with an arbitrary value of phase is always optimal for the cases of TFS, CS$\otimes$FS, as well as TMSVS.

Finally, we further investigate the maximal sensitivities given by CS$\otimes$PASVS and CS$\otimes$PSSVS. By analytically calculating the QFI, we present that both CS$\otimes$PASVS and CS$\otimes$PSSVS give a higher sensitivity than CS$\otimes$SVS under the constraint of the squeezing parameter, while the result is reversed under the
constraint of mean total photon number. More interestingly, both cases with photon addition and subtraction provide the same phase sensitivity for a fixed mean total photon number.

\section*{Appendix A: Derivation of Eqs.~(\ref{eq:two-QCRB-parity}) and (\ref{eq:single-QCRB-parity})
\label{sec:AA}}

  \makeatletter \renewcommand{\theequation}{A\arabic{equation}} \makeatother \setcounter{equation}{0}

In this appendix, we show the detailed derivation of Eq.~(\ref{eq:two-QCRB-parity}) under the assumption that the input state $\vert\psi_{{\rm in}}\rangle$ is a family of state associating with some parity symmetry, which requires one of input ports is injected by an even or odd state \cite{Liu2013PRA}. Here, we assume that the even or odd state is powered on mode $b$. With this, the expectation values of all operators in terms of odd-order moments of  $b$ in $\vert \psi_{\rm in}\rangle$ are vanishing, for instance, $\langle J_x\rangle=\langle J_y\rangle=0$. This is, in fact, a key step in the following derivations.
With Eq.~(\ref{eq:Jz-B}) and its squared counterpart
\begin{eqnarray}
(B^{\dagger}J_{z}B)^{2} & = & J_{z}^{2}\cos^{2}\tau+(J_{y}\cos\vartheta+J_{x}\sin\vartheta)^{2}\sin^{2}\tau\nonumber \\
 &  & +[J_{z}(J_{y}\cos\vartheta+J_{x}\sin\vartheta)+h.c.]\sin\tau\cos\tau,\nonumber \\
\end{eqnarray}
one gets the following equations
\begin{eqnarray}
\langle B^{\dagger}J_{z}B\rangle & = & \langle J_{z}\rangle\cos\tau,\\
\langle(B^{\dagger}J_{z}B)^{2}\rangle & = & \langle J_{z}^{2}\rangle\cos^{2}\tau+\langle J_{\perp,\vartheta}^{2}\rangle\sin^{2}\tau,\\
\langle J_{0}B^{\dagger}J_{z}B\rangle & = & \langle J_{0}J_{z}\rangle\cos\tau,
\end{eqnarray}
where we have set
\begin{eqnarray}
J_{\perp,\vartheta}&\equiv& J_{y}\cos\vartheta+J_{x}\sin\vartheta,
\end{eqnarray}
representing the component of angular momentum on the plane perpendicular to $z$ axis. Having these, we can express the QFI matrix elements given by Eqs.(\ref{eq:Fss}), (\ref{eq:Fsd}), and (\ref{eq:Fdd}) as follows:
\begin{eqnarray}
F_{ss} & = & 4\langle(\Delta J_{0})^{2}\rangle\label{eq:FssA}\\
F_{sd} & = & 4[\langle J_{0}J_{z}\rangle-\langle J_{0}\rangle\langle J_{z}\rangle]\cos\tau,\label{eq:FsdA}\\
F_{dd} & = & 4\langle(\Delta J_{z})^{2}\rangle\cos^{2}\tau+4\langle J_{\perp,\vartheta}^{2}\rangle\sin^{2}\tau.\label{eq:FddA}
\end{eqnarray}
With Eq.~(\ref{eq:FddA}), one can directly obtain the phase sensitivity bound provided by Eq.~(\ref{eq:single-QCRB-phid}) based on single-parameter estimation theory. Interestingly, we see that maximizing Eq.~(\ref{eq:FddA}) over parameters $\tau$ and $\vartheta$ is analogous to finding an optimal mean spin direction along which the variance of collective spin operator gets maximum in the atomic interferometry \cite{Ma2011PRA,Hyllus2012PRA,Pezze2012PRL}.

Reminding the notation of the Schwinger representation, one can further express Eqs.~(\ref{eq:FssA}), (\ref{eq:FsdA}) and (\ref{eq:FddA}) in terms of the mode operators $a$ and $b$ as below:
\begin{eqnarray}
F_{ss} & = & \langle(\Delta a^{\dagger}a)^{2}\rangle+\langle(\Delta b^{\dagger}b)^{2}\rangle+2(\langle a^{\dagger}ab^{\dagger}b\rangle\nonumber \\
 &  & -\langle a^{\dagger}a\rangle\langle b^{\dagger}b\rangle),\label{eq:FssM0}\\
F_{sd} & = & [\langle(\Delta a^{\dagger}a)^{2}\rangle-\langle(\Delta b^{\dagger}b)^{2}\rangle]\cos\tau,\label{eq:FsdM0}\\
F_{dd} & = & [\langle(\Delta a^{\dagger}a)^{2}\rangle+\langle(\Delta b^{\dagger}b)^{2}\rangle-2(\langle a^{\dagger}ab^{\dagger}b\rangle\nonumber \\
 &  & -\langle a^{\dagger}a\rangle\langle b^{\dagger}b\rangle)]\cos^{2}\tau+\mathfrak{F}\sin^{2}\tau,\label{eq:FddM0}
\end{eqnarray}
where we have set
\begin{eqnarray}
\mathfrak{F} & \equiv & 4\langle J_{\perp,\vartheta}^{2}\rangle\nonumber \\
 & = & -\langle\left(a^{\dagger}be^{i\vartheta}-ab^{\dagger}e^{-i\vartheta}\right)^{2}\rangle\nonumber \\
 & = & \langle a^{\dagger}a(b^{\dagger}b+1)\rangle+\langle(a^{\dagger}a+1)b^{\dagger}b\rangle-2{\rm Re}(e^{i2\vartheta}\langle a^{\dagger2}\rangle\langle b^{2}\rangle.\nonumber \\
\label{eq:maxQFI}
\end{eqnarray}
From Eqs.~(\ref{eq:FssA}), (\ref{eq:FsdA}), and (\ref{eq:FddA}) (or (\ref{eq:FssM0}), (\ref{eq:FsdM0}), and (\ref{eq:FddM0})), the sensitivity bounds depicted by Eqs.~(\ref{eq:two-QCRB-phid}) and (\ref{eq:single-QCRB-phid}) are identical for $\tau=\pi/2$ as
\begin{eqnarray}
\mathcal{V}_{2}(\hat{\phi}_{d})&=&\mathcal{V}_{1}(\hat{\phi}_{d})=(\upsilon\mathfrak{F})^{-1}.
\end{eqnarray}

If the input state be in a separable form $\vert\psi_{{\rm in}}\rangle=\vert\chi_{a}\rangle\vert\chi_{b}\rangle$, such that $\langle a^{\dagger}ab^{\dagger}b\rangle=\langle a^{\dagger}a\rangle\langle b^{\dagger}b\rangle$, we have the following identity:
\begin{eqnarray}
\langle(\Delta J_{0})^{2}\rangle&=&\langle(\Delta J_{z})^{2}\rangle=\frac{1}{4}\left[\langle(\Delta a^{\dagger}a)^{2}\rangle+\langle(\Delta b^{\dagger}b)^{2}\rangle\right].
\end{eqnarray}
Then Eqs.~(\ref{eq:FssM0}) and (\ref{eq:FddM0}) can be rewritten
as
\begin{eqnarray}
F_{ss} & = & \langle(\Delta a^{\dagger}a)^{2}\rangle+\langle(\Delta b^{\dagger}b)^{2}\rangle,\label{eq:FssM1}\\
F_{dd} & = & [\langle(\Delta a^{\dagger}a)^{2}\rangle+\langle(\Delta b^{\dagger}b)^{2}\rangle]\cos^{2}\tau+\mathfrak{F}\sin^{2}\tau.\label{eq:FddM1}
\end{eqnarray}
Submitting Eqs.~(\ref{eq:FssM1}), (\ref{eq:FsdM0}) and (\ref{eq:FddM1})
into Eq.~(\ref{eq:two-QCRB-phid}) finally yields

\end{multicols}
\begin{eqnarray}
\mathcal{V}_{2}(\hat{\phi}_{d}) & \geq & \frac{1}{\upsilon}\frac{\langle(\Delta a^{\dagger}a)^{2}\rangle+\langle(\Delta b^{\dagger}b)^{2}\rangle}{\left[\langle(\Delta a^{\dagger}a)^{2}\rangle+\langle(\Delta b^{\dagger}b)^{2}\rangle\right]F_{dd}-\left[\langle(\Delta a^{\dagger}a)^{2}\rangle-\langle(\Delta b^{\dagger}b)^{2}\rangle\right]^{2}\cos^{2}\tau}\nonumber \\
{\color{blue}} & = & \frac{1}{\upsilon}\frac{\langle(\Delta a^{\dagger}a)^{2}\rangle+\langle(\Delta b^{\dagger}b)^{2}\rangle}{\left[\langle(\Delta a^{\dagger}a)^{2}\rangle+\langle(\Delta b^{\dagger}b)^{2}\rangle\right]^{2}\cos^{2}\tau+\left[\langle(\Delta a^{\dagger}a)^{2}\rangle+\langle(\Delta b^{\dagger}b)^{2}\rangle\right]\mathfrak{F}\sin^{2}\tau-\left[\langle(\Delta a^{\dagger}a)^{2}\rangle-\langle(\Delta b^{\dagger}b)^{2}\rangle\right]^{2}\cos^{2}\tau}\nonumber \\
{\color{blue}} & = & \frac{1}{\upsilon}\frac{1}{\mathfrak{G}\cos^{2}\tau+\mathfrak{F}\sin^{2}\tau},\label{eq:bound}
\end{eqnarray}
\begin{multicols}{2}

where we have set
\begin{eqnarray}
\mathfrak{G} & \equiv & \frac{4\langle(\Delta a^{\dagger}a)^{2}\rangle\langle(\Delta b^{\dagger}b)^{2}\rangle}{\langle(\Delta a^{\dagger}a)^{2}\rangle+\langle(\Delta b^{\dagger}b)^{2}\rangle}.
\end{eqnarray}

\Acknowledgements{We thank Xiao-Ming Lu for the helpful discussions. This work was supported by the NSFC through Grants No. 11747161, 11974189, and a project funded by the Priority Academic Program Development of Jiangsu Higher Education Institutions. P.X. acknowledges the support from the NSFC through Grant No. 11847050 and the Natural Science Foundation of Jiangsu Province through Grant No. BK20180750.}

\begin{thebibliography}{99}

\bibitem{Caves1981PRD}C. M. Caves, Phys. Rev. D \textbf{23}, 1693 (1981).

\bibitem{Holland1993PRL}M. J. Holland and K. Burnett, Phys. Rev. Lett. \textbf{71}, 1355 (1993).

\bibitem{Luis2001PRA}A. Luis, Phys. Rev. A \textbf{64}, 054102 (2001).

\bibitem{Dorner2009PRL}U. Dorner, R. Demkowicz-Dobrzanski, B. J. Smith, J. S. Lundeen,
W. Wasilewski, K. Banaszek, and I. A. Walmsley, Phys. Rev. Lett. \textbf{102}, 040403 (2009).

\bibitem{Anisimov2010PRL}P. M. Anisimov, G. M. Raterman, A. Chiruvelli, W. N. Plick, S. D. Hu-
ver, H. Lee, and J. P. Dowling, Phys. Rev. Lett. \textbf{104},103602 (2010).


\bibitem{Joo2011PRL}J. Joo, W. J. Munro, and T. P. Spiller, Phys. Rev. Lett. \textbf{107}, 083601 (2011).

\bibitem{Pezze2013PRL}L. Pezze and A. Smerzi, Phys. Rev. Lett. \textbf{110}, 163604 (2013).

\bibitem{Tan2014PRA}Q. S. Tan, J. Q. Liao, X. G. Wang, and F. Nori, Phys. Rev. A \textbf{89}, 053822 (2014).

\bibitem{Lang2014PRA}M. D. Lang and C. M. Caves, Phys. Rev. A \textbf{90}, 025802 (2014).

\bibitem{Birrittella2014JOSAB}R. Birrittella and C. C. Gerry, J. Opt. Soc. Am. B \textbf{31}, 586 (2014).


\bibitem{Ouyang2016JOSAB}Y. Ouyang, S. Wang, and L. Zhang, J. Opt. Soc. Am. B \textbf{33}, 1373(2016).

\bibitem{Yuan2014}C. Yuan, K. Zhang, and W. Zhang, Scientia Sinica Inf. \textbf{44}, 345 (2014).

\bibitem{Liu2019}B. Liu, G. Li, Y. Che, J. Chen, and X. Wang, Sci. China Phys. Mech. Astron. \textbf{62}, 040301 (2019).

\bibitem{Zhang2016}X. Zhang and J. Ye, Nat. Sci. Rev. \textbf{3}, 189 (2016).

\bibitem{Nie2018}X. Nie, J. Huang, Z. Li, W. Zheng, C. Lee, X. Peng, and J. Du, Sci. Bull. \textbf{63}, 469 (2018).


\bibitem{Rarity1990PRL}J.G.Rarity, P.R.Tapster, E.Jakeman, T.Larchuk, R.A.Campos, M.C.
Teich, and B. E. A. Saleh, Phys. Rev. Lett. \textbf{65}, 1348 (1990).

\bibitem{Pezze2008PRL}L. Pezze and A. Smerzi, Phys. Rev. Lett. \textbf{100}, 073601 (2008).

\bibitem{Hofmann2009PRA}H. F. Hofmann, Phys. Rev. A \textbf{79}, 033822 (2009).

\bibitem{Krischek2011PRL}R. Krischek, C. Schwemmer, W. Wieczorek, H. Weinfurter, P. Hyllus,
L. Pezz¨¦, and A. Smerzi, Phys. Rev. Lett. \textbf{107}, 080504 (2011).

\bibitem{Zhong2017PRA}W. Zhong, Y. Huang, X. Wang, and S. L. Zhu, Phys. Rev. A \textbf{95}, 052304 (2017).


\bibitem{Seshadreesan2013PRA}K. P. Seshadreesan, S. Kim, J. P. Dowling, and H. Lee, Phys. Rev. A \textbf{87}, 043833 (2013).

\bibitem{Lang2013PRL}M. D. Lang and C. M. Caves, Phys. Rev. Lett. \textbf{111}, 173601 (2013).

\bibitem{Vidrighin2014NC}M. D. Vidrighin, G. Donati, M. G. Genoni, X. M. Jin, W. S. Kolthammer, M. S. Kim, A. Datta, M. Barbieri, and I. A. Walmsley, Nat.
Commun. \textbf{5}, 3532 (2014).

\bibitem{Liu2017PRA}P. Liu, P. Wang, W. Yang, G. R. Jin, and C. P. Sun, Phys. Rev. A \textbf{95}, 023824 (2017).

\bibitem{Helstrom1976Book}C. W. Helstrom, Quantum Detection and Estimation Theory (Academic, New York, 1976).


\bibitem{Holevo1982Book}A. S. Holevo, Probabilistic and Statistical Aspects of Quantum Theory (North-Holland, Amsterdam, 1982).

\bibitem{Braunstein1994PRL}S. L. Braunstein and C. M. Caves, Phys. Rev. Lett. \textbf{72}, 3439 (1994).


\bibitem{Long2018}G. L. Long, W. Qin, Z. Yang, and J. L. Li.,  Sci. China Phys. Mech. Astron. \textbf{61}, 030311 (2018).

\bibitem{Qin2019}W. Qin, A. Miranowicz, G. L. Long, J. Q. You, and F. Nori, npj Quantum Inf. \textbf{5}, 58 (2019).


\bibitem{Jarzyna2012PRA}M. Jarzyna and R. Demkowicz-Dobrzanski, Phys. Rev. A \textbf{85}, 011801(R) (2012).

\bibitem{Yurke1986PRA}B. Yurke, S. L. McCall, and J. R. Klauder, Phys. Rev. A \textbf{33}, 4033 (1986).


\bibitem{Lu2010PRA}X. M. Lu, X. G. Wang, and C. P. Sun, Phys. Rev. A \textbf{82}, 042103 (2010).


\bibitem{Zhong2013PRA}W. Zhong, Z. Sun, J. Ma, X. G. Wang, and F. Nori, Phys. Rev. A \textbf{87}, 022337 (2013).

\bibitem{Liu2019JPA}J. Liu, H. D. Yuan, X. M. Lu, and X. G. Wang, e-print arXiv:1907.08037.


\bibitem{Takeoka2017PRA}M. Takeoka, K. P. Seshadreesan, C. You, S. Izumi, and J. P. Dowling, Phys. Rev. A \textbf{96}, 052118 (2017).


\bibitem{Matsumoto2002JPA}K. Matsumoto, J. Phys. A: Math. Theor. \textbf{35}, 3111 (2002).

\bibitem{Pezze2017PRL}L. Pezze, M. A. Ciampini, N. Spagnolo, P. C. Humphreys, A. Datta,
I. A. Walmsley, M. Barbieri, F. Sciarrino, and A. Smerzi, Phys. Rev. Lett. \textbf{119}, 130504 (2017).


\bibitem{Baumgratz2016PRL}T. Baumgratz and A. Datta, Phys. Rev. Lett. \textbf{116}, 030801 (2016).

\bibitem{Gagatsos2016PRA}C. N. Gagatsos, D. Branford, and A. Datta, Phys. Rev. A \textbf{94}, 042342 (2016).

\bibitem{Molmer1997PRA}K. Molmer, Phys. Rev. A \textbf{55}, 3195 (1997).




\bibitem{Demkowicz-Dobrzanski2009PRA}R. Demkowicz-Dobrzanski, U. Dorner, B. J. Smith, J. S. Lundeen,
W. Wasilewski, K. Banaszek, and I. A. Walmsley, Phys. Rev. A \textbf{80}, 013825 (2009).



\bibitem{Trees2013Book}H. L. V. Trees, K. L. Bell and Z. Tian, Detection, Estimation, and
Modulation Theory, 2nd Edition, Part I. (Wiley, 2013). p.325.



\bibitem{Liu2013PRA}J. Liu, X. X. Jing, and X. G. Wang, Phys. Rev. A \textbf{88}, 042316 (2013).

\bibitem{Collaboration2011NP}L. S. Collaboration, Nat. Phys. \textbf{7}, 962 (2011).

\bibitem{Collaboration2013NP}L. S. Collaboration, Nat. Photon. \textbf{7}, 613 (2013).

\bibitem{Monras2006PRA}A. Monras, Phys. Rev. A \textbf{73}, 033821 (2006).



\bibitem{Genoni2011PRL}M. G. Genoni, S. Olivares, and M. G. A. Paris, Phys. Rev. Lett. \textbf{106}, 153603 (2011).

\bibitem{Gerry2004Book}C. Gerry and P. Knight, Introductory Quantum Optics (Cambridge University Press, Cambridge, 2004).

\bibitem{Hyllus2010PRL}P. Hyllus, L. Pezz¨¦, and A. Smerzi, Phys. Rev. Lett. \textbf{105}, 120501 (2010).

\bibitem{Pezze2015PRA}L. Pezze, P. Hyllus, and A. Smerzi, Phys. Rev. A \textbf{91}, 032103 (2015).

\bibitem{Kitagawa2006PRA}A. Kitagawa, M. Takeoka, M. Sasaki, and A. Chefles, Phys. Rev. A \textbf{73}, 042310 (2006).


\bibitem{Dakna1997PRA}M. Dakna, T. Anhut, T. Opatrny, L. Knoll, and D. G. Welsch, Phys. Rev. A \textbf{55}, 3184 (1997).

\bibitem{Kim2008JPB}M. S. Kim, J. Phys. B: At., Mol. Opt. Phys. \textbf{41}, 133001 (2008).

\bibitem{Wang2013EPJD}S. Wang, H. C. Yuan, and X. F. Xu, Eur. Phys. J. D \textbf{67}, 102 (2013).

\bibitem{Navarrete-Benlloch2012PRA}C. Navarrete-Benlloch, R. Garcia-Patr¨®n, J. H. Shapiro, and N. J. Cerf, Phys. Rev. A \textbf{86}, 012328 (2012).

\bibitem{Dakna1999PRA}M. Dakna, J. Clausen, L. Knoll, and D. G. Welsch, Phys. Rev. A \textbf{59}, 1658 (1999).



\bibitem{Fiurasek2005PRA}J. Fiurasek, R. Garcia-Patron, and N. J. Cerf, Phys. Rev. A \textbf{72}, 033822 (2005).

\bibitem{Biswas2007PRA}A. Biswas and G. S. Agarwal, Phys. Rev. A \textbf{75}, 032104 (2007).

\bibitem{Ourjoumtsev2007PRL}A. Ourjoumtsev, A. Dantan, R. Tualle-Brouri, and P. Grangier, Phys. Rev. Lett. \textbf{98}, 030502 (2007).

\bibitem{Duan2000PRL}L. M. Duan, G. Giedke, J. I. Cirac, and P. Zoller, Phys. Rev. Lett. \textbf{84}, 4002 (2000).

\bibitem{Eisert2004AP}J. Eisert, D. Browne, S. Scheel, and M. Plenio, Ann. Phys. \textbf{311}, 431 (2004).


\bibitem{Opatrny2000PRA}T. Opatrny, G. Kurizki, and D. G. Welsch, Phys. Rev. A \textbf{61}, 032302 (2000).

\bibitem{Cochrane2002PRA}P. T. Cochrane, T. C. Ralph, and G. J. Milburn, Phys. Rev. A \textbf{65}, 062306 (2002).

\bibitem{Olivares2003PRA}S. Olivares, M. G. A. Paris, and R. Bonifacio, Phys. Rev. A \textbf{67}, 032314 (2003).

\bibitem{Wenger2004PRL}J. Wenger, R. Tualle-Brouri, and P. Grangier, Phys. Rev. Lett. \textbf{92}, 153601 (2004).

\bibitem{Braun2014PRA}D. Braun, P. Jian, O. Pinel, and N. Treps, Phys. Rev. A \textbf{90}, 013821 (2014).



\bibitem{Wang2019OC}S. Wang, X. X. Xu, Y. J. Xu, and L. J. Zhang, Opt. Commum. \textbf{444}, 102 (2019).

\bibitem{Zavatta2004Sci}A. Zavatta, S. Viciani, and M. Bellini, Science \textbf{306}, 660 (2004).

\bibitem{Hong-Yi2006CTP}H. Y. Fan, X. G. Meng, and J. S. Wang, Commun. Theor. Phys. \textbf{46}, 845 (2006).

\bibitem{Hu2010JMO}L. Y. Hu and H. Y. Fan., J. Mod. Opt. \textbf{57}, 1344 (2010).

\bibitem{Meng2012JOS}X. G. Meng, Z. Wang, H. Y. Fan, and J. S. Wang, J. Opt. Soc. Am. B, \textbf{29}, 3141 (2012).



\bibitem{HONG-YI2019MPLA}H. Y. Fan, L. Y. Hu, and X. X. Xu, Mod. Phys. Lett. A \textbf{24}, 1597 (2019).

\bibitem{Barnett2018PRA}S. M. Barnett, G. Ferenczi, C. R. Gilson, and F. C. Speirits, Phys. Rev. A \textbf{98}, 013809 (2018).


\bibitem{Ma2011PRA}Jian Ma, Yi-xiao Huang, Xiaoguang Wang, and C. P. Sun, Phys. Rev. A \textbf{84}, 022302 (2011).


\bibitem{Hyllus2012PRA}P. Hyllus, W. Laskowski, R. Krischek, C. Schwemmer, W. Wieczorek, H. Weinfurter, L. Pezze, and A. Smerzi, Phys. Rev. A \textbf{85}, 022321 (2012).


\bibitem{Pezze2012PRL}L. Pezze and A. Smerzi, Phys. Rev. Lett. \textbf{102}, 100401 (2012).



\end{thebibliography}

\end{multicols}
\end{document}